\documentclass[12pt,oneside,a4paper,article]{revtex4}
\usepackage{graphicx}
\usepackage{dcolumn}
\usepackage{bm}
\usepackage{amsmath}
\usepackage[a4paper]{geometry}
\usepackage{microtype}
\usepackage{cleveref}
\usepackage{booktabs}
\usepackage{verbatim}

\begin{document}
\title{Isolating quantum coherence with pathway-selective coherent multi-dimensional spectroscopy}
\author{Jonathan O Tollerud, Christopher R Hall, Jeffrey A Davis}%
\affiliation{Centre for Atom Optics and Ultrafast Spectroscopy, Swinburne University of Technology, Hawthorn, 3122 Victoria, Australia}

\begin{abstract}
\textbf{ Coherent coupling between spatially separated systems has long been explored as a necessary requirement for quantum information and cryptography\cite{quantum}.  Recent discoveries suggest such phenomena appear in a much wider range of processes, including light-harvesting in photosynthesis\cite{EngelFleming,Scholes,Aspuru08}.  These discoveries have been facilitated by developments in coherent multi-dimensional spectroscopy (CMDS) \cite{brixner,LiCundiff,TurnerPhysChemChemPhys,Marcus} that allow interactions between different electronic states to be identified in crowded spectra.  For complex systems, however, spectral broadening and multiple overlapping peaks limit the ability to separate, identify and properly analyse all contributions\cite{HallNJP,DavisJCP2012}.  Here we demonstrate how pathway-selective CMDS can overcome these limitations to reveal, isolate and allow detailed analysis of weak coherent coupling between spatially separated excitons localised to different semiconductor quantum wells. Selective excitation of the coherence pathways, by spectrally shaping the laser pulses, provides access to previously hidden details and enables quantitative analysis that can facilitate precise and detailed understanding of interactions in this and other complex systems.
}
\end{abstract}

\maketitle

Coherent multi-dimensional spectroscopy for electronic transitions, much like equivalent techniques in infra-red (IR)\cite{ZanniHamm} and nuclear magnetic resonance (NMR) spectroscopy, utilises multiple pulses that excite and probe the sample during different time periods to quantify excited state dynamics and interactions between states.  In multi-dimensional NMR, this type of information  facilitates complete structure determination of complex molecules, such as proteins\cite{Ernstprotein}.  For electronic transitions, third order CMDS experiments have been used to explore energy transfer and relaxation dynamics, and more recently, to identify coherent coupling between excited states in light-harvesting complexes\cite{EngelFleming,Scholes}, conjugated polymers \cite{Collini} and well-separated semiconductor nanostructures\cite{langbeinNatureP}.  

In these experiments three phase-locked pulses generate a signal with phase and amplitude that is measured by a heterodyne detection scheme. Varying the delays between pulses results in three time periods ($t_1$, $t_2$, $t_3$) and three corresponding frequency domains ($\omega_1, \omega_2, \omega_3$), as described in Methods and Supplementary Information.  To analyse these data, 2D spectra that correlate the absorption energy ($\hbar\omega_1$) and the emission energy ($\hbar\omega_3$) for different values of the delay $t_2$, are typically presented.  

Coherent coupling can be identified in such experiments in the form of a coherent superposition of states, which generates peaks with phase that oscillates as a function of $t_2$. Alternatively, Fourier transforming the data with respect to $t_2$ shifts these features along $\hbar\omega_2$ by an amount equal to the energy difference between the coupled states\cite{DavisJCP2012,HallNJP,Cundiff3D,TurnerReview}.  In simple systems these coherence pathways can thus be separated from other signal pathways that involve population relaxation, energy transfer, ground state bleach and excited state absorption. In complex systems, however, such as light-harvesting complexes from photosynthetic organisms, numerous states and spectral broadening lead to overlapping peaks that can be difficult, if not impossible, to identify or separate.
Additionally, for systems where many-body effects are important, such as semiconductor nanostructures, excitation of transitions at one energy can alter the signal detected despite playing no direct role in its generation\cite{LiCundiff}, thus further complicating the interpretation.

The origin of these limitations is the same broad spectral bandwidth that makes 2D spectroscopy so useful. On the one hand, the ability to explore multiple pathways simultaneously can speed-up data acquisition, and the analysis of 2D peak shapes can provide more information than is otherwise accessible. On the other hand, if the many different pathways cannot be separated, these advantages are lost.  Several important and useful approaches to separate different pathways in broadband experiments have been established, \cite{Read2007,TurnerNature,TurnerPhysChemChemPhys,NelsonSelectiveEnhancement} 
yet there often remain contributions that cannot be isolated, leading to difficulties and uncertainty in the analysis. 

Two-colour four-wave experiments that selectively excite and probe specific coherence pathways have recently shown some advantages over broadband CMDS\cite{RichardsJPCL,Womick}.
What has been lacking, however, is the phase stability between pulses that allows multi-dimensional spectra to be obtained, and with it the ability to analyse peak-shapes and fully correlate the relative contributions from different pathways.

We have combined the selectivity achieved in these approaches with the phase stability required for CMDS, allowing us to perform both broadband and pathway-selective experiments that can be quantitatively compared. 
To achieve this we utilise a CMDS apparatus based on a pulse shaper (as pioneered by Nelson \textit{et al.}\cite{TurnerReview}) to delay and compress each of the beams and independently shape their spectral amplitudes. 
Different pathways can then be selected by controlling the spectrum of each pulse, while the intrinsic phase stability is maintained, allowing multi-dimensional spectra from only the desired pathways to be measured.  

\begin{figure}
     \includegraphics[width=.5\textwidth]{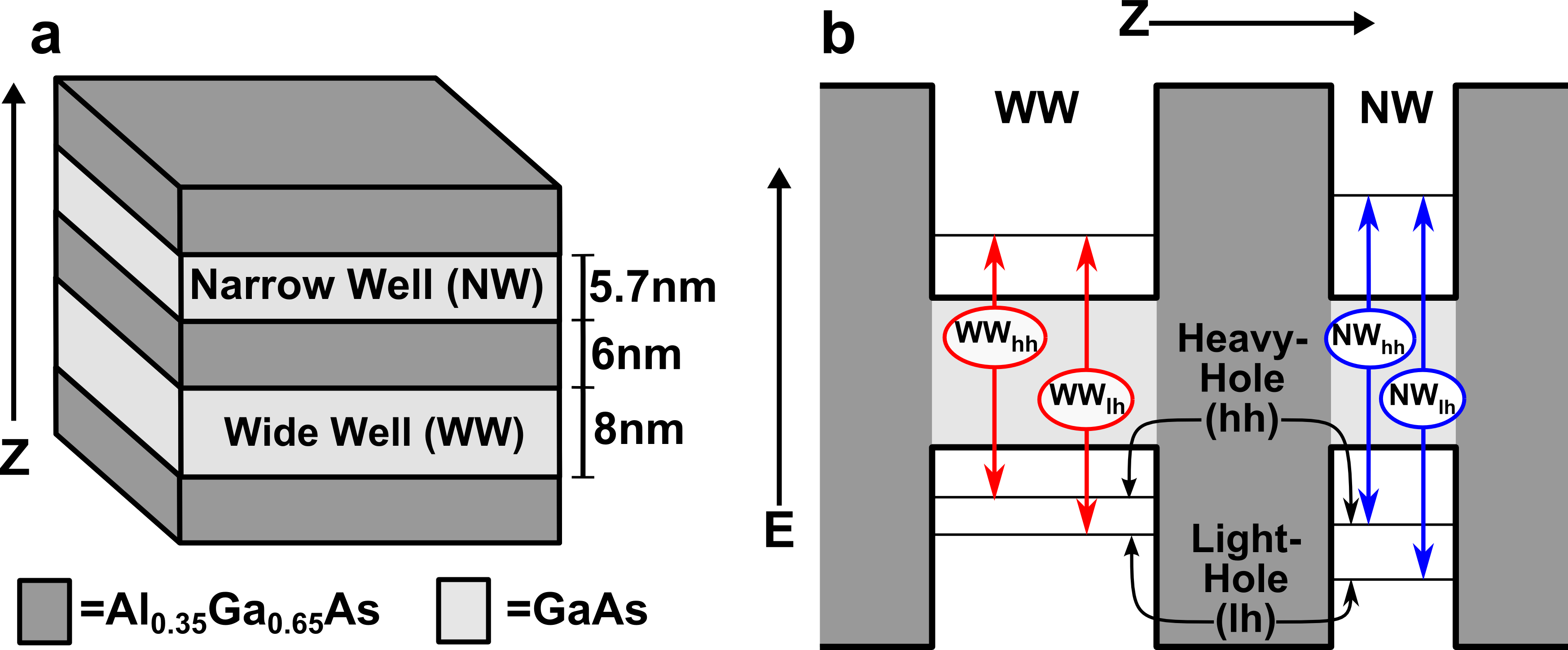}
 \caption{(a) The asymmetric double quantum well structure shows layers of GaAs 5.7~nm and 8~nm thick (the QWs) separated by a 6~nm layer of Al$_{0.35}$Ga$_{0.65}$As (the barrier). The profile of the potential perpendicular to the layers is shown in (b), with four bright transitions between electron and hole states localised to each well, as indicated and labelled.}
 \label{QWs}
 \end{figure}

We utilise this approach to reveal and explore coherent coupling between excitons localised to semiconductor quantum wells (QWs) separated by a 6~nm barrier, as depicted in Fig.~\ref{QWs}. The different widths of the two QWs lead to different transition energies and the possibility of downhill energy and/or charge transfer between wells \cite{Devaud,Leo,DirectedTransfer}.  This type of system has been explored extensively for potential device applications \cite{ADQWdevices} and as a tunable template to explore fundamental energy transfer processes\cite{HallNJP,CundiffADQW}. When the barrier between wells is low and/or narrow substantial coupling between the wells leads to hybridised wavefunctions and a significant role for coherent quantum effects in energy and charge transfer.  For higher, wider barriers, where the electron and hole wavefunctions are localised to single QWs, the role and nature of quantum coupling between wells is less clear. In the case of the 6~nm barriers used here each of the wavefunctions are well localised to one of the wells (as shown in Supplementary Information) and coherent coupling between wells is expected to be weak, if present at all.

\begin{figure*}
\includegraphics[width=0.5\textwidth]{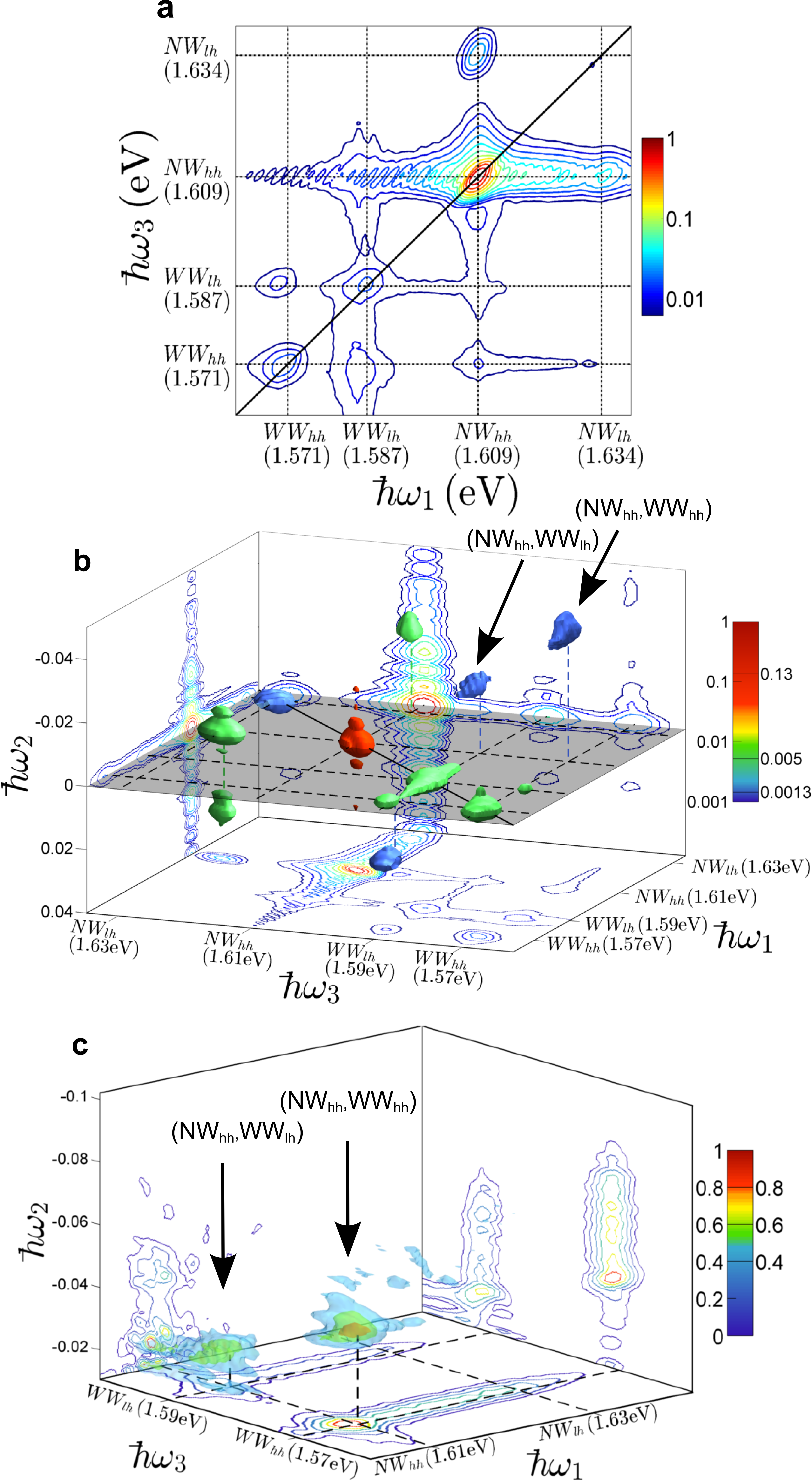}
\caption{Experimental results from broadband CMDS experiments. (a) shows the 2D spectrum at $t_2=0$.  The dashed lines provide a guide to indicate the energy for each of the 4 bright transitions.  (b) the 3D spectrum is represented as a series of isosurfaces, with the projections in each directions. The isosurfaces shown are plotted only for certain regions to minimise the complexity of the spectrum and highlight specific peaks. The full isosurfaces at each level are shown in Supplementary Information.  The strongest peaks occur at $\hbar\omega_2=0$, indicating population pathways, while peaks away from this plane represent coherent superpositions.  The inter-well coherence peaks are shown in more detail in (c) where the 3D spectrum and projections confirm the location and origin of these peaks.}
\label{broadband}
\end{figure*}

Broadband CMDS (Fig.~\ref{broadband}) is able to identify coherent coupling between the narrow well heavy-hole $NW_{hh}$ exciton and each of wide well $WW$ excitons through the presence of cross-peaks that are shifted along $\hbar\omega_2$ by amounts equal to the energy differences between the coupled exciton states. These inter-well cross-peaks are, however, almost three orders of magnitude weaker than the strongest peaks and as a result sit on top of a large noisy background that is proportional to the total signal at each $\hbar\omega_3$ value (see Supplementary Information for details).  The 2D spectrum (a) shows contributions from pathways involving each of the four excitons indicated in Fig.~\ref{QWs}(b), but is dominated by the $NW_{hh}$ diagonal peak; while the cross peaks combine contributions from both population and coherence pathways. The 3D spectrum (b) separates these contributions and shows the majority of the signal to be at $\hbar\omega_2=0$ and therefore due to population pathways. Coherences involving heavy-hole and light-hole excitons localised to the same well are the next strongest contributions, while the two inter-well coherence peaks are barely above the noise.

Figure \ref{broadband}(c) shows these inter-well coherence peaks in isolation, but due to the poor signal to noise little analysis beyond identifying their presence is possible. In contrast, the intra-well coherences, which are much stronger and well above the noise, demonstrate peak shapes that are elongated in the diagonal direction, indicating correlated inhomogeneous broadening (see Supplementary Information). 


\begin{figure*}
\includegraphics[width=0.6\textwidth]{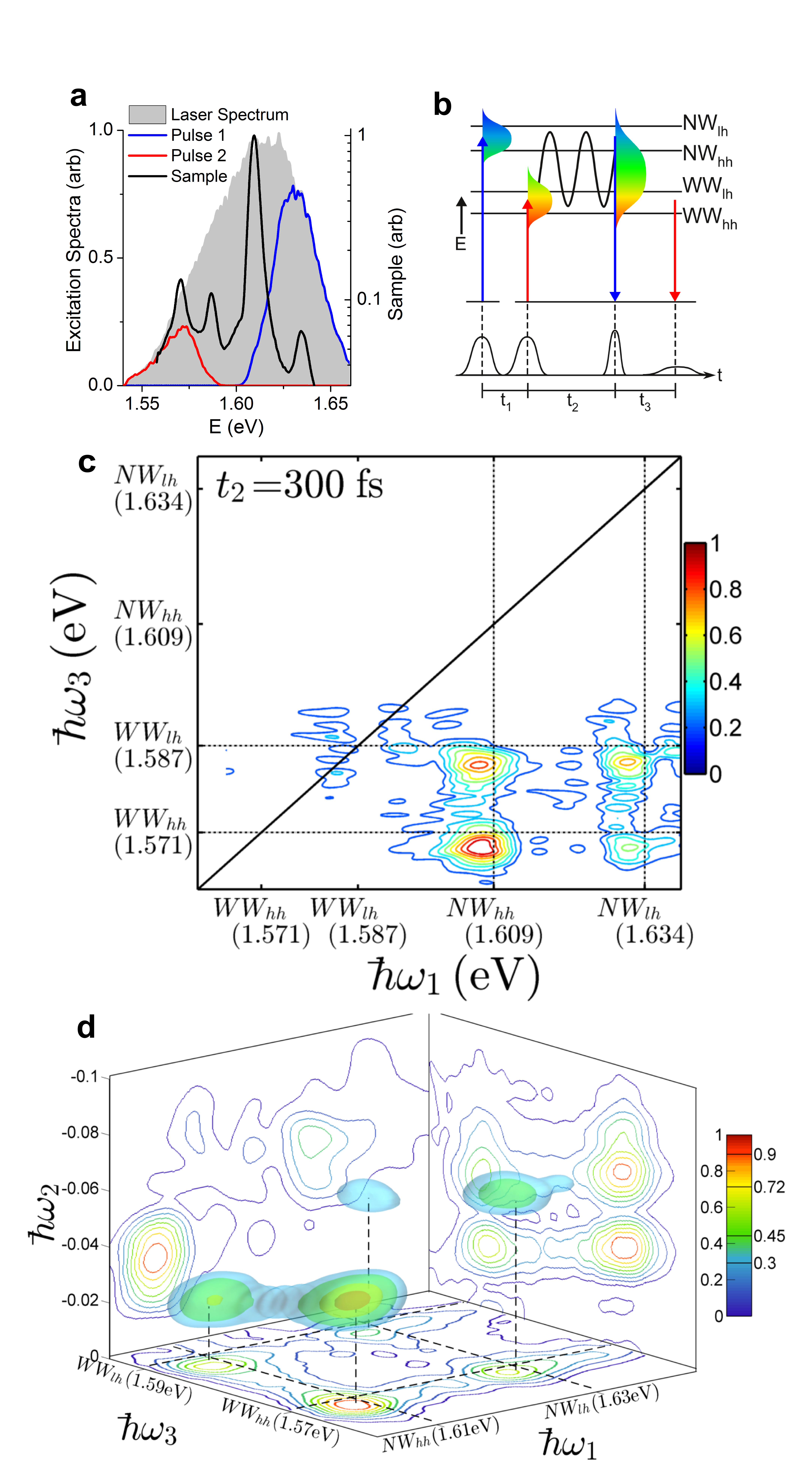}
\caption{(a) the broadband laser spectrum and the spectrum for the first two pulses in the pathway selective experiment are shown on top of the spectrum from the quantum well sample.   The pulse sequence and states excited in the coherence specific experiment are shown in (b). The resultant 2D spectrum at $t_2=300$~fs, (c), shows only the four inter-well cross-peaks.  The 3D spectrum (d) confirms that these peaks arise entirely from inter-well coherence pathways.  The separation of the four peaks in three dimensions and enhanced signal to noise allows further quantitative and peak shape analysis.}
\label{twocolour}
\end{figure*}

To further examine the inter-well coherence peaks the pathways that lead to these signals were selectively excited using the pulse sequence shown in Fig.~\ref{twocolour}(b). With the first pulse resonant only with NW excitons and the second resonant only with WW excitons, all population and single well coherence pathways will be excluded.

A 2D spectrum using this pulse sequence is shown in Fig.~\ref{twocolour}(c). By comparison to the broadband 2D spectrum (Fig.~\ref{broadband}(a)) it can be seen that all single well processes are suppressed and the only signal is in the region of the inter-well cross peaks. There are four inter-well peaks present, the two identified in Fig.~\ref{broadband} and two additional peaks at ($NW_{lh}$,$WW_{hh}$) and ($NW_{lh}$,$WW_{lh}$).  The 3D spectrum in Fig.~\ref{twocolour}(d) shows each of the peaks to be well-resolved in all three frequency domains and confirms that these four peaks are all due to inter-well coherent superpositions.  The absence of signal at $\hbar\omega_2 = 0$ further confirms that the coherent superposition pathways are indeed being excited in isolation. 

Isolating the coherence pathways has not only identified two additional coherent signals (and hence inter-well coupling between two additional pairs of excitons), but also enhanced the signal to noise, allowing further detailed analysis of the shape, location and magnitude of each peak.

The peak shapes of these inter-well coherences show no sign of being elongated along the diagonal, in contrast to the intra-well coherence peaks, indicating that inhomogeneous broadening of excitons localised in different wells is uncorrelated.  This is explained by the major source of broadening being local fluctuations in the well widths, which are not correlated \cite{DavisJCP2012}.    The uncorrelated nature of the broadening is further confirmed by analysis that shows the peak width in $\hbar\omega_2$ to be equal to the sum of the widths along $\hbar\omega_1$ and $\hbar\omega_3$, as shown in Supplementary Information. In contrast, fully correlated broadening would ensure the width in $\hbar\omega_2$ is equal to the homogeneous linewidth and thus narrower than the inhomogeneous linewidths measured along $\hbar\omega_1$ and $\hbar\omega_2$.
  
The isolation of these signal pathways reveals that each peak is shifted to lower energy by 1-2~meV in both $\hbar\omega_1$ and $\hbar\omega_3$ compared to the diagonal peak locations in the broadband measurements.  We speculate that the origin of this shift is a reduction of the excitation induced shift (EIS) as a result of narrowing the excitation spectra and reducing the overall density of carriers. Previous work in GaAs QWs has shown that EIS is important and can be identified by careful peak shape analysis of the real part of the signal\cite{LiCundiff,Turner2012}. To confirm that the shift we see is due to reduced EIS, further experiments to determine the global phase and isolate the real part of the susceptibility are required.

Further quantitative analysis, as described in the Supplementary Information, shows that the strongest of these coherence peaks is more than three orders of magnitude smaller than the diagonal peaks measured in the broadband spectra.  Following corrections for spectral amplitudes at each transition energy the $(NW_{hh},WW_{hh})$ coherence peak is shown to be the strongest, followed by $(NW_{hh},WW_{lh})$, $(NW_{lh},WW_{hh})$ and $(NW_{lh},WW_{lh})$.  With further experiments and analysis it should be possible to determine precisely all transition dipole moments  and the coupling strength between each of the spatially separated excitons. This type of quantitative analysis with dynamic range exceeding three orders of magnitude presents the opportunity to extend recent demonstrations of quantum state and process tomography in simple systems\cite{AspuruQPT,Cundiff3D,NelsonQPT} to larger and more complex systems.


In conclusion, we have devised a pathway specific CMDS experiment that combines many of the benefits of CMDS with an ability to selectively excite specific quantum pathways. We utilise these capabilities to unambiguously reveal coherent coupling between excitons localised to quantum wells separated by 6~nm.  The coherence peaks are well-isolated, allowing identification of two peaks not previously seen as well as peak shape analysis and quantitative comparisons that are not possible with the equivalent data from broadband CMDS. 

The ability to isolate these coherences, and indeed any specific signal pathway, can provide significant insight into the interactions and dynamics in a range of complex systems.  In photosynthetic light harvesting complexes, for example, this type of approach has the potential to resolve important questions regarding the nature and role of quantum effects in efficient energy transfer.

\section*{Methods}
We utilised a Titanium:Sapphire oscillator to produce transform limited $\sim$45~fs pulses centered at 785~nm (as confirmed by X-FROG measurements) at a repetition rate of 97~MHz. The CMDS experimental apparatus utilised two spatial light modulators (Boulder Nonlinear 512 nematic SLM) in an arrangement similar to Turner et al. \cite{TurnerReview}.  The first SLM is used as a Fourier beam shaper to split the incident beam into four beams in a boxcars geometry, necessary for excitation and heterodyne detection. These are relayed through a 4F imaging system (with all beams incident on the same optics to provide intrinsic phase stability) to a pulse shaper based on the second SLM. Each beam is spectrally dispersed horizontally and separated from the other beams vertically on the SLM.  A spectral phase is applied to each beam independently to compensate for any chirp and apply the specified delay (a linear phase gradient in frequency corresponds to a shift in the time domain).  In addition, a vertical grating is applied to diffract the beams down.  This allows the time delayed beams to be picked off from the incident beams and allows amplitude control of each of the beams by varying the depth of the vertical grating, which facilitates spectral shaping \cite{Vaughan}.  The delayed beams are then imaged to the sample, where they overlap and excite a third order polarization that radiates in momentum conserving directions. At the sample position each of the three excitation beams have average power of 2.8~mW and are focussed to a $100\mu m$ diameter spot. The four-wave mixing signal detected is collinear with the local oscillator and focussed into a spectrometer where spectral interferometry allows the amplitude and phase of the signal to be determined.
An eight-step phase cycling procedure is used to minimise noise and scatter from the excitation beams and maximise the signal.   

To generate a 3D spectrum the delay $t_1$ was scanned in 10~fs steps for fixed values of the delay $t_3$, which was varied in 15~fs steps. For all of the data presented here only the rephasing contribution ($t_1>0$)is shown. A rotating frame of reference was used, with the carrier frequency set to 795~nm.  This ensures that the phase at 795~nm does not change as the delays are varied and reduces the sampling requirements for complete determination of the electric fields.  From the spectral interferograms the amplitude and phase are determined and the data Fourier transformed with respect to $t_1$ and $t_2$.

In the coherence pathway specific experiment, spectral amplitude shaping was used to tailor the excitation spectrum of the first two pulses so that they were centred on different transitions with very little spectral overlap.  The spectral amplitude masks used are shown in Supplementary Information and were chosen to give spectral amplitudes that were close to Gaussian.  The flat spectral phase leads to transform limited pulses and the Gaussian spectrum ensures good temporal profiles, also shown in Supplementary Information.  The average powers of these pulses are then reduced to 1.0~mW and 0.3~mW for the first two pulses, respectively. 
For the third excitation beam and the local oscillator the full laser spectrum was used.
This pulse scheme drives only processes that involve coherent superpositions between states excited by the first two pulses respectively. 

A major benefit of this experimental setup is the flexibility. The pulse sequence utilised here is one of many possible combinations which can be designed to match the temporal and spectral requirements of the sample and select any given pathway. The intrinsic ability to perform a series of such pathway-selective CMDS experiments, alongside broadband CMDS, with no changes to the optical setup will allow quantitative comparisons that facilitate precise and detailed understanding of interactions in complex systems. 

The GaAs/AlGaAs double quantum well sample was grown by Metal-Organic Chemical Vapour Deposition (MOCVD) and throughout the experiments was cooled to 20K in a closed-cycle circulating cryostat.

\section{Acknowledgments}
The authors would like to thank Prof C. Jagadish and Prof H.H. Tan at the Australian National University and The National Fabrication Facility for the quantum well sample. This work was supported by the Australian Research Council (FT120100587).


\section{Supplementary Information}
For Supplementary Information or any questions please email jdavis@swin.edu.au


\end{document}